\documentclass[fleqn,usenatbib]{mnras}

\usepackage{newtxtext,newtxmath}
\usepackage[T1]{fontenc}
\usepackage{ae,aecompl}

\usepackage{graphicx}	% Including figure files
\usepackage{amsmath}	% Advanced maths commands
\usepackage{amssymb}	% Extra maths symbols

\title[First discovery of an UCWD benchmark]{First discovery of an ultra-cool white dwarf benchmark\\ in common proper motion with an M dwarf}

\author[Lam et al.]{M C Lam$^{1}$\thanks{E-mail: c.y.lam@ljmu.ac.uk},
N C Hambly$^{2}$\thanks{E-mail: nch@roe.ac.uk},
N Lodieu$^{3,4}$,
S Blouin$^{5}$,
E J Harvey$^{1}$,
R J Smith$^{1}$,
\newauthor M C G{\'a}lvez-Ortiz$^{6}$ and
Z H Zhang$^{7}$
\\
$^{1}$Astrophysics Research Institute, Liverpool John Moores University, IC2, LSP, 146 Brownlow Hill, Liverpool L3~5RF, UK\\ 
$^{2}$Institute for Astronomy, University of Edinburgh, Royal Observatory of Edinburgh, Blackford Hill, Edinburgh EH9~3HJ, UK\\
$^{3}$Instituto de Astrof\'isica de Canarias~(IAC), Calle V\'ia L\'actea s/n, E-38200 La Laguna, Tenerife, Spain\\
$^{4}$Departamento de Astrof\'isica, Universidad de La Laguna (ULL), E-38206 La Laguna, Tenerife, Spain \\
$^{5}$Los Alamos National Laboratory, P.O. Box 1663, Mail Stop P365, Los Alamos, NM 87545, USA \\
$^{6}$Suffolk University, Madrid Campus, C/Valle de la Vi\~na 3, 28003, Madrid, Spain \\
$^{7}$School of Astronomy and Space Science, Key Laboratory of Ministry of Education, Nanjing University, Nanjing 210023, China
}

\date{Accepted 2020 February 25. Received 2020 January 15; in original form 2019 December 11}

\pubyear{2019}

\begin{document}
\label{firstpage}
\pagerange{\pageref{firstpage}--\pageref{lastpage}}
\maketitle

\begin{abstract}
Ultra-cool white dwarfs are among the oldest stellar remnants in the Universe. Their efficient gravitational settling and low effective temperatures are responsible for the smooth spectra they exhibit. For that reason, it is not possible to derive their radial velocities or to find the chemistry of the progenitors. The best that can be done is to infer such properties from associated sources, which are coeval. The simplest form of such a system is a common proper motion pair where one star is an evolved degenerate and the other a main sequence star. In this work, we present the discovery of the first of such a system, the M dwarf LHS 6328 and the ultra-cool white dwarf PSO J1801+625, from the Pan-STARRS 1 3$\pi$ survey and the {\it Gaia} Data Release 2. Follow-up spectra were collected covering a usable wavelength range of $3750$ to $24500$\,{\AA}. Their spectra show that the white dwarf has an effective temperature of $3550$\,K and surface gravity of $\log{g}=7.45\pm0.13$ or $\log{g}=7.49\pm0.13$ for a CO or He core, respectively, when compared against synthetic spectra of ultra-cool white dwarf atmosphere models. The system has slightly subsolar metallicity with $-0.25<[$Fe$/$H$]<0.0$, and a spatial velocity of $\left(U, V, W\right) = \left(-114.26\pm0.24, 222.94\pm0.60, 10.25\pm0.34\right)$\,km\,s$^{-1}$, the first radial velocity and metallicity measurements of an ultra-cool white dwarf. This makes it the first and only benchmark of its kind to date.
\end{abstract}


%%%%%%%%%%%%%%%%%%%%%%%%%%%%%%%%%%%%%%%%%%%%%%%%%%

%%%%%%%%%%%%%%%%% BODY OF PAPER %%%%%%%%%%%%%%%%%%

\section{Introduction}
White dwarfs~(WDs) are the final stage of stellar evolution of
main sequence~(MS) stars with zero age MS~(ZAMS) mass
less than $8\mathcal{M}_{\sun}$. Since this mass range
encompasses the vast majority of stars in the Galaxy, these
degenerate remnants are the most common final product of stellar
evolution, thus providing a good sample to study the history of stellar
evolution and star formation in the Galaxy. In this state, there is
little nuclear burning to replenish the energy they radiate away.
As a consequence, the luminosity and temperature decrease
monotonically with time. The electron degenerate nature means that
a WD with a typical mass of $0.6\mathcal{M}_{\sun}$ has a similar
size to the Earth, giving rise to their high densities, low
luminosities, and large surface gravities. Typically, it takes less
than $\mathcal{O}$$(10^4)$ yr for metals to sink beneath the
photosphere~\citep{2009A&A...498..517K}, leaving only hydrogen
and/or helium in the atmosphere. This leads to the lack of
metallic features in the atmosphere. WDs with carbon and oxygen
features are likely to have atypical evolutionary
pathways~\citep{2016Sci...352...67K}. The surface
temperatures~(T$_{\mathrm{eff}}$) are $\sim$$10^{5}$\,K among the
hottest WDs and can be as cool as $\sim$$3000$\,K at the faint
end. When they cool below $\sim$$5000$\,K, the hydrogen
and helium lines in the optical also disappear, leaving a featureless
spectrum.

Ultra-cool white dwarfs~(UCWDs) are the population of featureless
cool WDs~(spectral type DC) that exhibit collisionally induced
absorption~(CIA) features. This effect sets in at around
T$_{\mathrm{eff}}=4000$\,K. This temperature limit changes depending
on the chemistry of the atmosphere. Recent works suggest a distinction
between UCWD and IR faint cool 
WDs~\citep{2012MNRAS.423L.132K, 2015MNRAS.449.3966G}. These cool and
ultra-cool WDs have neutral colours and very low luminosities, so they
are very hard to find and, therefore, study. Hence, they have been of
interest to the science community for a number of years; however, we still
know little about them. One of the keys is to better understand the
structural and chemical evolution of the atmosphere as well as the core.
Consequentially, they have significant impact in the understanding
of the age and evolution of all Galactic components: through inversion
of the colour-magnitude diagram~\citep{2002A&A...390..917V,
2006A&A...459..783C}, or WD luminosity
functions~\citep{2013MNRAS.434.1549R, 2014ApJ...791...92T}.

Common proper motion~(CPM) binaries are gravitationally bound pairs 
formed when their birth cluster dissolves~\citep{2010MNRAS.404..721M,
2011MNRAS.415.1179M, 2011ASPC..451....9K}. Their wide physical
separations mean that they have weak binding energy -- the orbits can
easily be disrupted; hence, they are assumed to be coeval and
have evolved independently. A large sample of such binaries can be used
to probe the Galactic gravitational potential~\citep{2004ApJ...601..289C,
2009MNRAS.400.2128Q, 2014ApJ...790..159M, 2017A&A...600A..59C}. CPM
systems containing a WD and a burning star provide the advantage
that the age of the system can be estimated from WD cooling,
while the progenitor chemistry and the radial velocity of the system can
be extracted from the co-moving counterpart. Various works have studied
their kinematics~\citep{2001AJ....121..503S}, contribution to the dark
matter of the Galaxy~\citep{2002AJ....124.1118S}, stellar
chronology~\citep{2006ApJ...638..446M, 2019ApJ...870....9F}, the
initial-final mass relation (IFMR; \citealt{2008A&A...477..213C,
2012ApJ...746..144Z, 2015ApJ...815...63A, 2018ApJ...860L..17E}). As the
final products of post-asymptotic giant branch~(AGB) stars, the mass
distribution for WDs with companions is particularly useful to constrain
wide binary orbital evolution as a function of spectral type~\citep{2001AJ....121..503S}.

In this work, we present the first glimpse of the UCWD
PSO~J180153.685+625419.450\footnote{Due to its high proper motion, in
Pan-STARRS DR1, this object is identified as two sources, the other source ID is
PSO~J180153.689+625419.758 with $19$ epochs. The designation we have chosen,
PSO~J180153.685+625419.450, has $71$ epochs of measurements.}. It is the first UCWD
found in CPM with a hydrogen-burning low-mass star
where the chemistry of the progenitor system can be found. Section~$2$
describes the identification of this CPM pair and the follow-up
observation programme is described in Section~$3$. The analyses of the
M~dwarf and the UCWD are presented in Section~$4$. At the end, in
Section~$5$, we discuss the possible origin of the degenerate component.

\section{Identification of the CPM pair}
\begin{figure}
    \centering
    \includegraphics[width=\columnwidth]{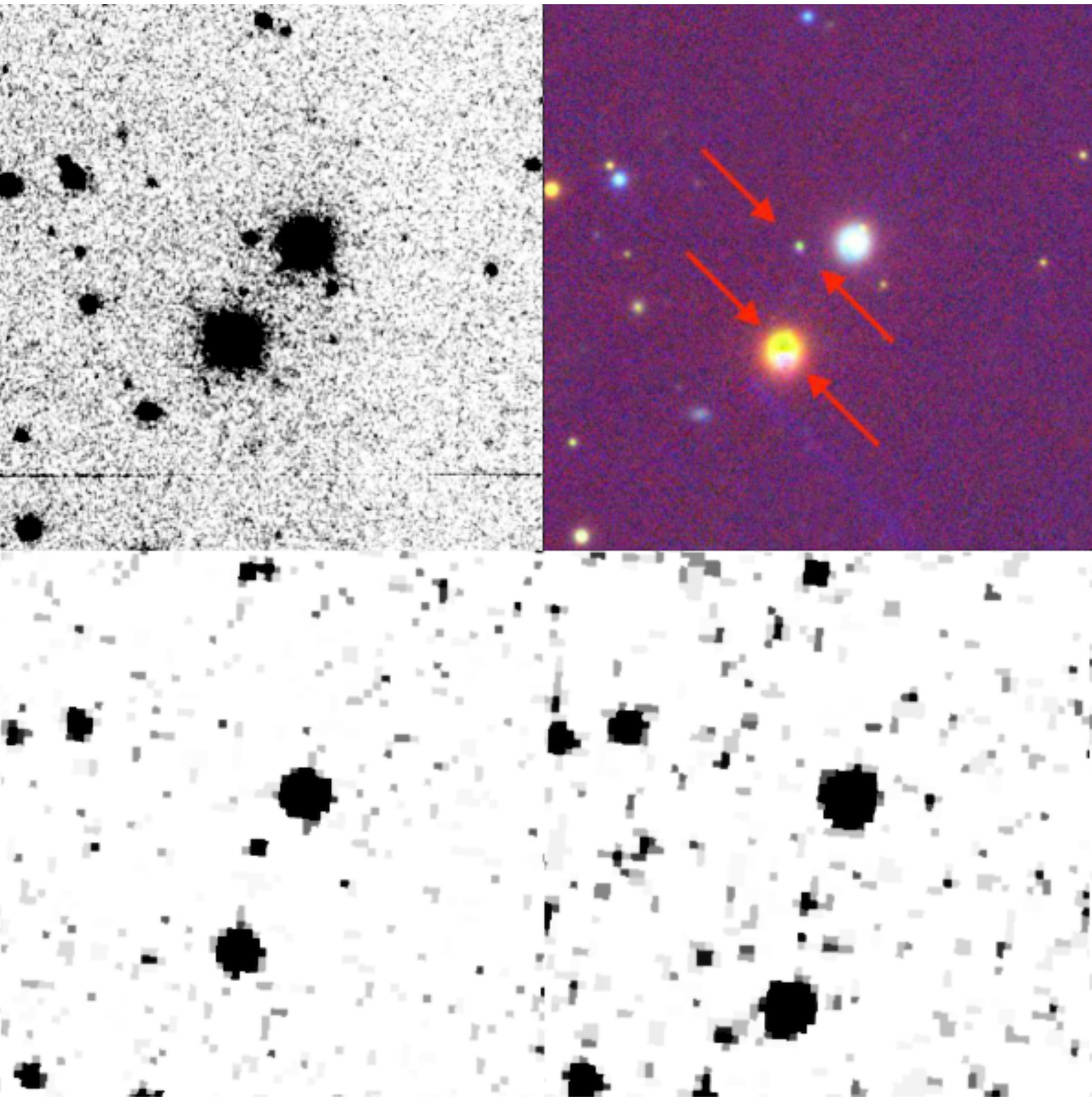}
    \caption{Clockwise from top left: LT/IO:O $r$-band image (2019); Pan-STARRS1 $g$/$i$/$y$ false-colour image (2012); the First Palomar Sky Survey plate image (1952); and Guide Star Catalogue 1 plate image (1983) of the PSO~J1801+625~(faint object centred at the Pan-STARRS 1 image) and LHS~6328~(bright).}
    \label{fig:finder_chart}
\end{figure}

From the WD catalogue~\citep[hereafter L19]{2019MNRAS.482..715L} selected 
large proper motion sources from the Pan-STARRS~1 $3\pi$
Survey~\citep{2016arXiv161205560C, 2016arXiv161205242M, 2016arXiv161205240M,
2016arXiv161205244M, 2016arXiv161205245W}, where there are a number of
candidate UCWDs. When parallaxes from the {\it Gaia} DR2~(GDR2;
\citealt{2016A&A...595A...1G,2018A&A...616A...1G})
became available, PSO~J1801+625 was immediately confirmed to have unusually
low luminosity at M$_{\mathrm{bol}}>16$\,mag. A proper motion and parallax
cross match was performed and the WD was found to be in CPM
with LHS\,6328, a well catalogued M dwarf with a wealth of
photometric information but never studied in detail given its unremarkable
properties~(see Fig.~\ref{fig:finder_chart} for the false colour image from
Pan-STARRS~1\footnote{\url{https://ps1images.stsci.edu/cgi-bin/ps1cutouts}}).
Further adding to the interest of this system is the tangential velocity
from  GDR2 (corresponding to the M dwarf distance) which suggests a likely
thick disc origin, $v_\mathrm{tan} = 4.74047 \times \mu \times D = 125.46$\,km\,s$^{-1}$.
We list the parameters of the system in Table~\ref{tab:observed_properties}
and ~\ref{tab:fitted_properties}. Using photometry from Pan-STARRS~1, and
the GDR2 parallax of the M dwarf, the best fit photometric
effective temperature and atmospheric composition can be found by
maximising the extended likelihood function from L19,
\begin{equation}
\mathcal{L} = -\frac{1}{2} \sum_{i} \left\{ \left[ \dfrac{(\mathrm{mag}_{i} - \mathrm{DM}) - \mathrm{model}_{i}}{\sigma_{i}} \right]^{2} + \ln{(2\pi \sigma_{i}^{2})} \right\}
\end{equation}
where index $i$ denotes the filter, mag is the observed magnitude, DM is
the distance modulus, model is the synthetic photometry in absolute
magnitude, and
\begin{equation}
\sigma_{i}^{2} = \sigma_{\mathrm{mag},i}^{2} + \left[ \frac{5 \log{(e)}}{\varpi} \right]^{2} \sigma_{\varpi}^{2}
\end{equation}
is the combined uncertainty in magnitude $\sigma_{\mathrm{mag}, i}$ and
in parallax $\sigma_{\varpi}$, where $\varpi$ is the parallax.

For the low mass main sequence star the GDR2 Apsis-Priam effective
temperature is $3675$\,K, placing it in the range where objects are
consistently overestimated~\citep{2018A&A...616A...8A}. The mean offset
is about $300$\,K, thus classifying this object as a M1-M3
dwarf~\citep{2013A&A...556A..15R,2018A&A...610A..19R}. The angular
separation of the system is $21.30''$, combining with the distance of the
M dwarf from \citet{2018AJ....156...58B}, the projected physical separation
of the system is $1554.1$\,AU, which is a common value for CPM wide
binaries~\citep{2015ApJ...815...63A, 2018ApJ...860L..17E}.

This system was not reported by a recent large-scale CPM work using GDR2
data because of the large photometric uncertainty of the faint WD
component~\citep{2018ApJ...860L..17E}. However, in our case, we have
reliable photometry from Pan-STARRS~1 complemented with the high-quality
astrometry from GDR2. With all of this supporting evidence we conducted a
follow-up campaign described in the following Section.

\begin{table}
  \centering
  \begin{tabular}{c c c}
    Parameter                             & LHS~6328             & PSO~J1801+625 \\\hline\hline
    $^{(a)}$R.A. ($\alpha$)    & $18$:$01$:$54.228$   & $18$:$01$:$53.691$ \\
    $^{(a)}$Dec. ($\delta$)        & $+62$:$53$:$59.432$  & $+62$:$54$:$20.414$\\
    $^{(a)}$\verb+source_id+      & \scriptsize{$2159435171891858176$}  & \scriptsize{$2159435274971074048$} \\
    $^{(a)}\mu_{\alpha}/$\,mas\,yr$^{-1}$ & $10.210 \pm 0.046$   & $10.498 \pm 1.122$\\
    $^{(a)}\mu_{\delta}/$\,mas\,yr$^{-1}$ & $362.712 \pm 0.055$  & $360.240 \pm 0.964$\\
    $^{(a)}\varpi/$\,mas                  & $13.6769 \pm 0.0258$ & $14.7551 \pm 0.5680$\\
    $^{(a)}$ ruwe                         & $1.152$              & $1.043$\\
    $^{(b)}$Distance / pc                 & $72.962^{+0.139}_{-0.138}$ & $67.737^{+2.733}_{-2.532}$\\
    %Distance Modulus / mag                & $4.315$              & $4.154$ \\
    $^{(c)}G$ / \,mag                   & $14.490 \pm 0.0003$  & $20.105 \pm 0.007$\\
    $^{(c)}G_{\mathrm{BP}}/$\,mag       & $15.632 \pm 0.004$   & $20.575 \pm 0.081$\\
    $^{(c)}G_{\mathrm{RP}}/$\,mag       & $13.430 \pm 0.001$   & $19.145 \pm 0.053$\\
    $^{(d)}B_{\mathrm{APASS}}/$\,mag    & $17.011 \pm 0.011$   & --- \\
    $^{(d)}V_{\mathrm{APASS}}/$\,mag    & $15.314 \pm 0.062$   & --- \\
    $^{(e)}g_{\mathrm{P1}}/$\,mag       & $16.016 \pm 0.003$   & $21.259 \pm 0.028$\\
    $^{(e)}r_{\mathrm{P1}}/$\,mag       & $14.769 \pm 0.003$   & $20.019 \pm 0.018$\\
    $^{(e)}i_{\mathrm{P1}}/$\,mag       & $13.875 \pm 0.007$   & $19.517 \pm 0.024$\\
    $^{(e)}z_{\mathrm{P1}}/$\,mag       & $13.419 \pm 0.007$   & $19.311 \pm 0.035$\\
    $^{(e)}y_{\mathrm{P1}}/$\,mag       & $13.181 \pm 0.002$   & $19.268 \pm 0.027$\\
    $^{(f)}J_{\mathrm{2MASS}}/$\,mag    & $12.031 \pm 0.021$   & --- \\
    $^{(f)}H_{\mathrm{2MASS}}/$\,mag    & $11.508 \pm 0.022$   & --- \\
    $^{(f)}K_{\mathrm{2MASS}}/$\,mag    & $11.292 \pm 0.020$   & --- \\
    $^{(g)}W_{1}/$\,mag                 & $11.147 \pm 0.023$   & --- \\
    $^{(g)}W_{2}/$\,mag                 & $10.979 \pm 0.020$   & --- \\
    $^{(g)}W_{3}/$\,mag                 & $10.896 \pm 0.051$   & --- \\
    $^{(g)}W_{4}/$\,mag                 & $>9.380$             & --- \\
    \hline\hline
  \end{tabular}
  \caption{This table lists all the relevant archival data available in the public domain. All photometry is reported in AB magnitude except for the 2MASS magnitudes that are reported in the Vega system. $^{(a)}$~Positions are in equinox J2000.0 and epoch J2015.5 as reported in GDR2. $^{(b)}$~Distance taken from \citet{2018AJ....156...58B}. $^{(c)}$~GDR2. $^{(d)}$~AAVSO Photometric All Sky Survey~(APASS)~\citep{2015AAS...22533616H}. $^{(e)}$~Pan-STARRS 1 3$\pi$ Survey. $^{(f)}$~2MASS~\citep{2006AJ....131.1163S}. $^{(g)}$~ALLWISE Data Release~\citep{2014yCat.2328....0C}.} 
  \label{tab:observed_properties}
\end{table}

\begin{table}
  \centering
  \begin{tabular}{c c}
    Parameter                             & Value \\\hline\hline
    DA Photometric$^{(a)}$ T$_{\mathrm{eff}}/$\,K & $3645^{+71}_{56}$\\
    DA Photometric$^{(a)}$ Distance / pc          & $58.8^{+1.6}_{-1.3}$\\
    DA Bolometric Magnitude$^{(a)}$               & $16.25^{+0.07}_{-0.08}$\\
    DB Photometric$^{(a)}$ T$_{\mathrm{eff}}/$\,K & $3158^{+25}_{-91}$\\
    DB Photometric$^{(a)}$ Distance / pc          & $44.2^{+1.6}_{-2.0}$\\
    DA Bolometric Magnitude$^{(a)}$               & $16.89^{+0.12}_{-0.10}$\\
    \hline
    Photometric$^{(b)}$ T$_{\mathrm{eff}}/$\,K & $4264^{+7}_{-6}$\\
    Photometric$^{(b)}$ log($\mathcal{M}_{\mathrm{He}}/\mathcal{M}_{\mathrm{H}}$) & $-1.61^{+0.07}_{-0.05}$\\
    \hline
    DA Photometric$^{(c)}$ T$_{\mathrm{eff}}/$\,K & $3959$\\
    DA Photometric$^{(c)}$ log(g) & $8.045$\\
    DA Photometric$^{(c)}$ mass / $\mathcal{M}_{\sun}$ & $0.605$\\
    DB Photometric$^{(c)}$ T$_{\mathrm{eff}}/$\,K & $4121$\\
    DB Photometric$^{(c)}$ log(g) & $8.017$\\
    DB Photometric$^{(c)}$ mass / $\mathcal{M}_{\sun}$ & $0.578$\\
    \hline\hline
  \end{tabular}
  \caption{Derived properties~(apart from the distance) of the UCWD. $^{(a)}$ Using Pan-STARRS~1 photometry fitting DA and DB models with fixed $\log(g)=8.0$ as reported in L19. $^{(b)}$ Using Pan-STARRS~1 photometry and GDR2 parallax with fitting mixed atmosphere model with fixed $\log(g)=8.0$. $^{(c)}$ Values reported by \citet{2019MNRAS.482.4570G} using GDR2 photometry and parallax.}
  \label{tab:fitted_properties}
\end{table}

\section{Follow-up Observations}
We collected additional photometric points and low resolution spectra 
with the $2$\,m Liverpool Telescope~(LT; \citealp{2004SPIE.5489..679S}), 
the $4$\,m William Herschel Telescope~(WHT) and the $10$\,m Gran Telescopio
Canarias, covering a wavelength range of $3750 - 25000$\,\AA\..

\subsection{LT}
Under the proposal \verb+JL18A03+, low-resolution spectra of the UCWD and
the M dwarf were collected with the SPRAT spectrograph in red-optimised 
mode~\citep{2014SPIE.9147E..8HP}. For the UCWD, five $20$-minute exposures
were collected on each of the dark nights on 2018 August 6 and 7; for the
M dwarf, five $2$\,min spectra were collected on 2018 August 10.

SPRAT has a resolving power of R$\sim$$300$, covering a usable wavelength
range of $4000 - 8000$\,{\AA}~(blue line in Fig.~\ref{fig:UCWD_spectrum}).
We re-extracted the 1D spectra from the in-house calibrated 2D
spectra~(\verb+LSS_NONSS+ of the Level 2 products). Flux
calibration was done using observations of the standard Hilt\,102 taken on the
same nights as the science observations. The
spectra were then median-summed to give our final spectra, delivering a
signal-to-noise ratio in the range $5-10$ at different wavelengths for the
UCWD and about $200$ for the M dwarf. Some residual telluric features could
not be completely removed.

In addition to the SPRAT spectra, three $20$\,min medium-resolution spectra
were collected with FRODOSpec at the high resolution setting~(R$\sim$$5300$)
covering a wavelength range of $5900 - 8000$\,{\AA}. Due to the high
sensitivity of the detector, we applied cosmic ray rejection
with L.A.Cosmic~\citep{2012ascl.soft07005V}, before processing with the
in-house automated pipeline~\citep{2012ASPC..461..517B}. The final spectrum
is the weighted sum of the spectra from the $12$ brightest fibres.

Additional photometric points were collected in Sloan $g$, $r$, $i$ and $z$
bands in the optical with IO:O and $H$ band with IO:I in the near-infrared,
which has a red cut-off at $17000$\,\AA. The $H$-band data was slightly
affected by the saturation residual mark from the neighbouring background
star. All data were collected in photometric and dark condition.

\subsection{WHT}
Under the service proposal \verb+SW2018A35+, two half-hour science frames
were collected on each night during the 2018 August 1 and 2 under bright
condition with a moon phase of around $70$ per cent. The spectra were collected
with WHT/ISIS using the R300B and R136R blue and red gratings centered at
$4499$ and $8651${\AA}, respectively. The Intermediate-dispersion
Spectrograph and Imaging System~(ISIS) is a high-efficiency double-arm
spectrograph with a long-slit of up to $4$ arcmin mounted on the WHT. The
ISIS spectrograph is equipped with a blue and red detector made of
$2048\times4096$ pixels whose sizes are $13.5$ and $15$ microns,
respectively. The choice of this setup was made to maximize the
wavelength coverage, i.e.\ where the LT/SPRAT and WHT/ACAM do not have
coverage.

Standard long-slit single object data reduction was performed using
\textsc{iraf}~\citep{1986SPIE..627..733T,1993ASPC...52..173T}. 2D Spectra
were median averaged separately for each night, before the two 1D spectra
were optimally extracted. The final spectrum was the weighted average of
the wavelength and flux calibrated spectra using the \textsc{SpectRes}
package\footnote{\url{https://github.com/ACCarnall/spectres}}~\citep{2017arXiv170505165C}.
Standard observations of G93-48 were taken immediately before and after the
observation on the respective nights, allowing for good removal of the
telluric features, where only small residuals remain. The absolute flux
calibration was done by comparing the integrated flux in the range of
$4250-5250$\,{\AA} between the absolute calibrated SPRAT spectrum and the
ISIS blue arm. Both the red and blue ISIS spectra were then corrected by
the same factor. 

Three exposures of $600$\,s were obtained for the WD with the
Auxiliary-port CAMera (ACAM) instrument mounted on the WHT on the night of
2019 July 29 between UT\,=\,21h37 and 22h10, under the program
095-WHT10/19A (PI G{\'a}lvez-Ortiz; observer Lodieu). One single exposure
of 300s was collected for the M dwarf companion starting at UT\,=\,21h27.
ACAM offers low-resolution (R$\sim$$400$) optical ($3500-9400$\,{\AA})
spectroscopy with a slit of 1 arcsec and the second-order blocking filter
GG495.  Arc lamps of Cu$+$Ne and tungsten flat field were collected
immediately after at the position of the targets. A spectroscopic standard,
Ross\,640~\citep[DZA5.5][]{1993PASP..105..761W}, was observed at the
beginning of the night to correct for the response of the detector and
calibrate the flux of our target.

The data were reduced using
\textsc{iraf}~\citep{1986SPIE..627..733T,1993ASPC...52..173T}. Bias
subtraction and flat field correction were undertaken with the exposures
obtained immediately following the science frames. The spectrum was
extracted optimally choosing the aperture and the sky regions carefully.
The final WHT/ACAM spectrum of the WD is displayed in green in
Fig.~\ref{fig:UCWD_spectrum} while the spectrum of the M dwarf is shown in
red in Fig.~\ref{fig:template_fitting}.

\subsection{Gran Telescopio Canarias}
Photometric and spectroscopic information was obtained with the Espectrografo Multiobjeto Infra-Rojo~\citep[EMIR][]{2007RMxAC..29...12G,2016SPIE.9908E..1JG} on the 10.4-m Gran Telescopio de Canarias (GTC) at the Observatorio del Roque de los Muchachos on the island of La Palma, under the programme GTC11-19A (PI: Lodieu). EMIR provides imaging and spectroscopic capabilities at near-infrared ($0.9-2.5$\,microns)  wavelengths. It is equipped with a $2048\times2048$ Teledyne HAWAII-2 HgCdTe detector with a pixel size of $0.2$\,arcsec per pixel, offering a $6.64\times6.64$ arcmin field-of-view.

Photometry was collected for the wide binary system on 2019 July 22 in
service mode with a seeing of $0.6$\,arcsec, during bright time, and clear
sky as part of programme GTC11\_19A (PI Lodieu). Total exposure times of
$175$, $175$, $315$, and $1743$\,s were set for the $Y$, $J$, $H$, and
$K_{s}$ passbands with on-source individual integrations of $5$\,s and a
pattern of seven dithers to avoid saturation of the brightest stars next
to the common proper motion pair. Spectroscopy was conducted in service
mode as part of the same GTC programme with a seeing of $0.7$ arcsec,
bright time, and clear skies. Four and eight on-source integrations of
$360$\,s in an ABBA pattern were collected for the WD on 10 August 2019
with the $YJ$ and $HK$ grisms of EMIR covering the $0.899-1.331$ and
$1.454-2.428$ micron intervals at a spectral resolution of about $500$ with
a slit width of $1.0$\,arcsec. A spectro-photometric standard star of
spectral type B9III, HIP84021~\citep{1970MmRAS..72..233H}, was observed
immediately after the target in both grisms.

The EMIR photometry and spectroscopy were reduced with the \verb+pyemir+
pipeline \footnote{\url{https://pyemir.readthedocs.io/en/latest/}}. For 
the photometric data reduction, a median-filtered flat field was
subtracted from each individual raw image before re-projecting the images
with the {\tt{reproject}} utility in Python. Afterwards, the images were
astrometrically calibrated with Gaia sources present in the EMIR field of
view. The parallactic angle was set for the photometric and spectroscopic
observations. Ten stars in the field were used to compute the ensemble
photometry. The $H$-band photometry agrees with the independent photometry
from the LT. The photometric results is are listed in
Table~\ref{tab:followup_observed_properties} in the AB magnitude system.

For the spectroscopic data reduction, the pyemir pipeline implements a
flat field correction, applies a wavelength calibration, and finds the
offsets between the A and B positions of the spectra along the slit. The
extraction of the spectra of the target and standard in each grism was
conducted manually under \textsc{iraf}, for selecting of the optimal
apertures. The spectrum of the target was then divided by the spectrum of
the standard and multiplied by a blackbody of the same temperature,
smoothed to the EMIR resolution. The final spectra were down-sampled to
$18.4$\,{\AA}\,pix$^{-1}$, half the wavelength resolution of the optical
spectra. The near-infrared spectrum of the WD is plotted in 
Figure~\ref{fig:UCWD_spectrum}.

\begin{table}
  \centering
  \begin{tabular}{c c c}
    Filter                   & Magnitude \\\hline\hline
    g$_{\mathrm{LT}}/$\,mag  & $21.12\pm0.07$\\
    r$_{\mathrm{LT}}/$\,mag  & $20.15\pm0.04$\\
    i$_{\mathrm{LT}}/$\,mag  & $19.51\pm0.03$\\
    z$_{\mathrm{LT}}/$\,mag  & $19.26\pm0.03$\\
    J$_{\mathrm{GTC}}/$\,mag & $19.45\pm0.02$\\
    H$_{\mathrm{LT}}/$\,mag  & $19.90\pm0.05$\\
    H$_{\mathrm{GTC}}/$\,mag & $19.96\pm0.03$\\
    K$_{\mathrm{GTC}}/$\,mag & $20.40\pm0.05$\\
    \hline\hline
  \end{tabular}
  \caption{Follow-up observations of the UCWD in AB magnitude.}
  \label{tab:followup_observed_properties}
\end{table}

\section{Analysis}
\subsection{LHS 6328}
\begin{figure}
    \centering
    \includegraphics[width=\columnwidth]{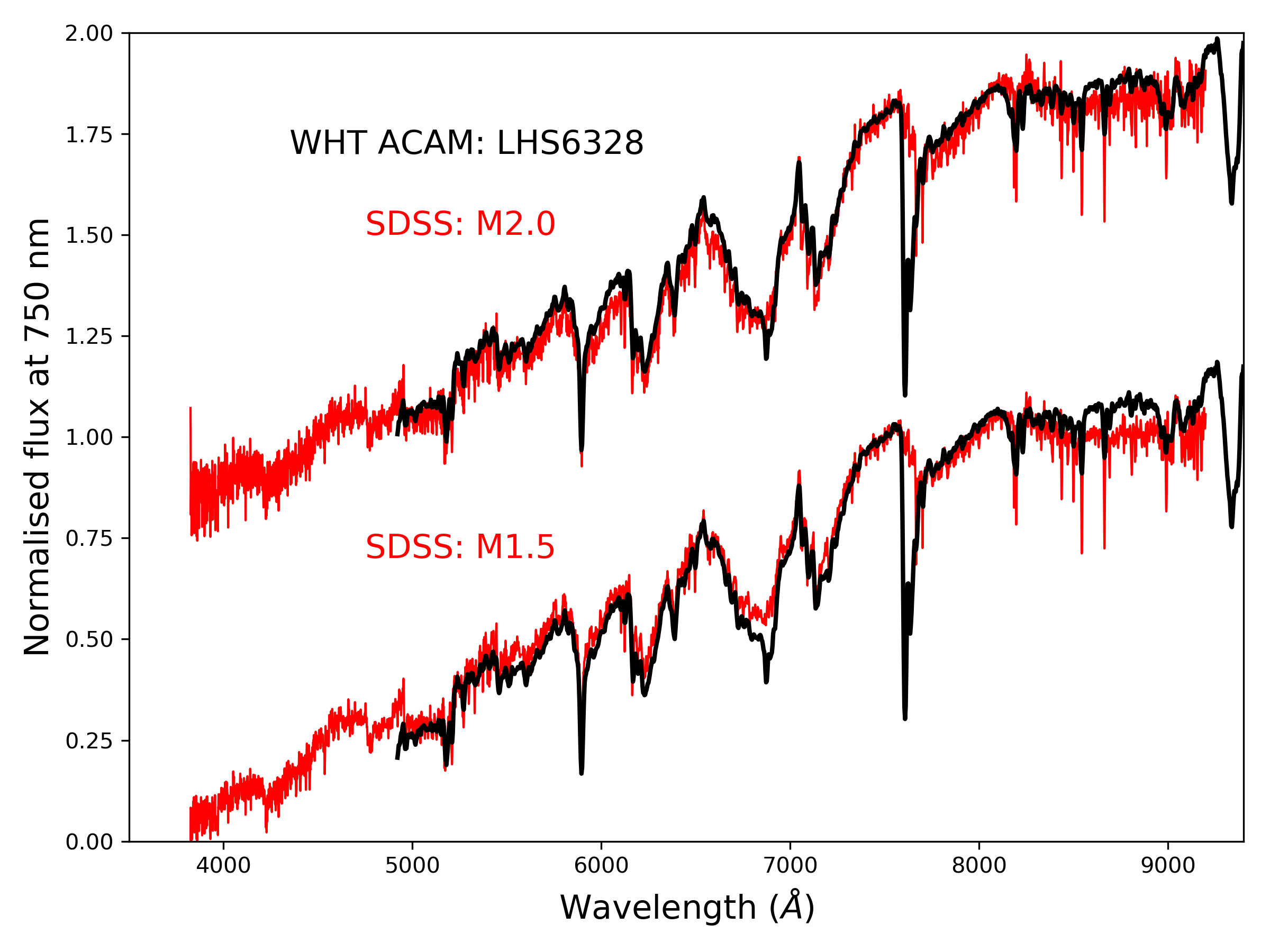}
    \caption{Template~(red) fit of the ACAM spectrum~(black) that gives the best-fit solution between type M$1.5$ and M$2$ with solar metallicity. Therefore, we assign it as a type M$1.75\pm0.25$ dwarf. The spectra are normalised at $7500${\AA}. The ACAM spectrum is not corrected for telluric absorption bands while the Sloan spectra~(red) are; hence there is a clear discrepancy at around $7600$\,{\AA} due to atmospheric O$_2$.}
    \label{fig:template_fitting}
\end{figure}

The ACAM spectrum of the MS star was compared against the low mass star
template spectra from SDSS~\citep{2007AJ....133..531B}. Using the FRODOSpec
spectra, we find that the metallicity indicators,
$\zeta_{\mathrm{TiO/CaH}}$, from \citet{2007ApJ...669.1235L} and
\citet{2019ApJS..240...31Z} are 0.92 and 0.86, respectively. Both values
suggest a slightly metal-poor M dwarf. The TiO5 spectral indexes show that
they are dM1.70~\citep{1995AJ....110.1838R},
dM1.94~\citep{1997AJ....113..806G} and dM1.77~\citep{2019ApJS..240...31Z},
consistent with the best-fit solution by eye as a type M$1.75\pm0.25$ dwarf
with roughly solar-metallicity~(Fig.~\ref{fig:template_fitting}).

\begin{figure*}
    \centering
    \includegraphics[width=\textwidth]{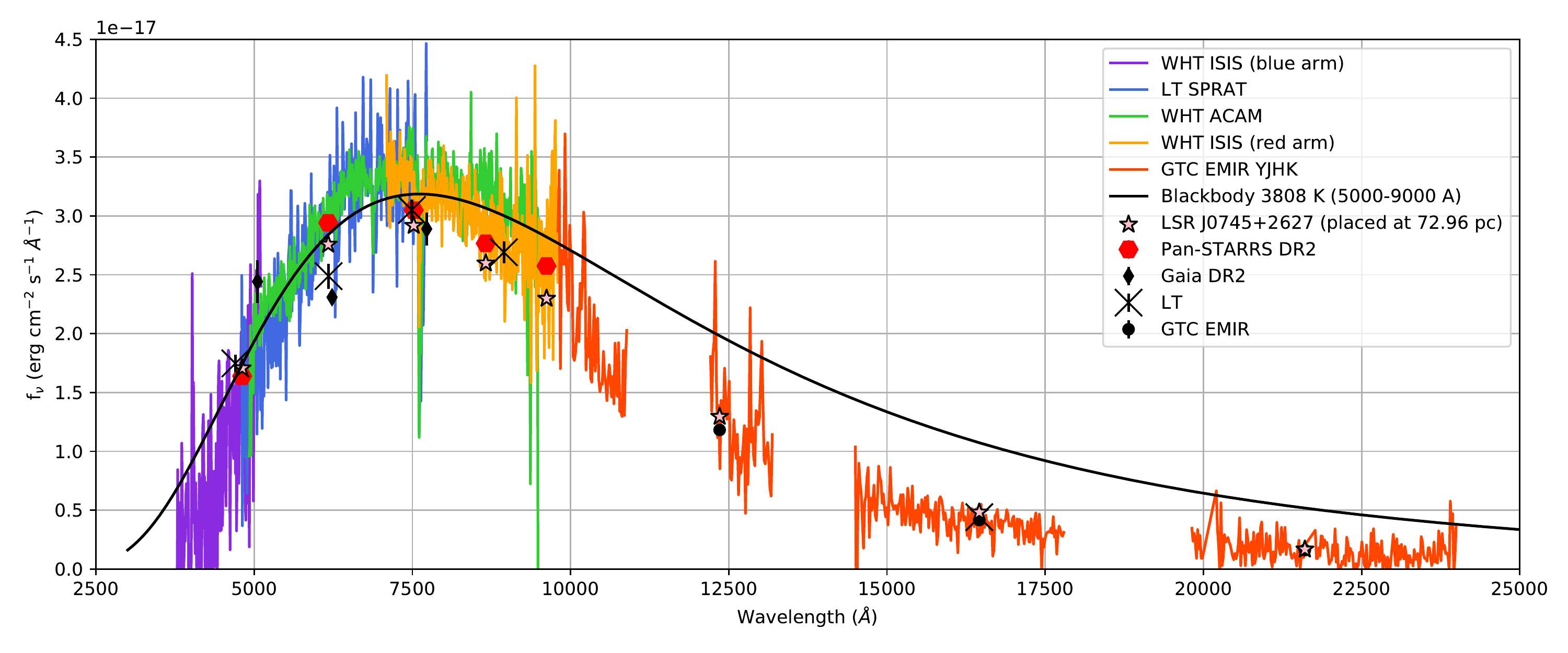}
    \caption{The spectra collected from WHT/ISIS blue and red arm~(purple/orange), LT/SPRAT~(blue), WHT/ACAM~(green) and GTC/EMIR~(red) overplotted with a black-body spectrum at $3808$\,K~(black, $\chi^2$-fitted over the wavelength range of $5000-9000$\,\AA). All optical spectra are re-sampled to $9.2$\,\AA\,per pixel, while the EMIR spectra are smoothed to $18.4$\,\AA\, per pixel. Only the usable ranges of the respective spectra are shown. The spectrum is smooth and does not show any absorption lines apart from the residual sky lines, consistent with the spectrum from a cool WD atmosphere. Pan-STARRS 1 $g/r/i/z/y$~(left to right), {\it Gaia} DR2 $G_{\mathrm{BP}}/G/G_{\mathrm{RP}}$ and LT/IO:O and IO:I $g/r/i/z/H$ photometry~(left to right) were converted to flux using the pivot wavelengths of the respective filters. All the overlapping ranges of the spectra agree well with each other.}
    \label{fig:UCWD_spectrum}
\end{figure*}

The radial velocity of the system was found with eight lines that can be
identified by eye are measured from the median of the three FRODOSpec spectra,
and compared against the
BT-Settl\footnote{\url{https://phoenix.ens-lyon.fr/Grids/BT-Settl}}
model atmosphere of a M dwarf at $3500$\,K~\citep{2012RSPTA.370.2765A,
2015A&A...577A..42B}. The model spectrum is refraction-corrected with an
air density of $0.733$\,amg based on the observing conditions. The median
and standard deviation, computed from the product of $1.46$ and the median
absolute deviation~(MAD), of the eight solutions are 
$v_{\mathrm{r}}=-1.19 \pm 0.69 $\,km\,s$^{-1}$. The full list can be found
in the Appendix on Table~\ref{tab:rv_lines}. Using the \textsc{AstroPy}
\verb+coordinates+ and \verb+time+ packages, the barycentric correction was
found to be $-1.34$\,km\,s$^{-1}$ yielding a barycentric radial velocity of
$-2.53\pm0.69$\,km\,s$^{-1}$. By drawing $10000$ samples from the Gaussian
distributions of the proper motions, parallax and radial velocities with
their associated uncertainties as the standard deviation, the
\verb+transform_to+ method gives us the Galactic velocity\footnote{Detailed descriptions can be found
at \url{https://docs.astropy.org/en/stable/api/astropy.coordinates.Galactocentric.html}}
$\left(U, V, W\right) = \left(-114.26\pm0.24, 222.94\pm0.60, 10.25\pm0.34\right)$\,km\,s$^{-1}$, we use the default Solar spatial motion 
with the \texttt{AstroPy} package: $\left(U_{\sun}, V_{\sun}, W_{\sun}\right) = (11.1, 232.24, 7.25)$\,km\,s$^{-1}$.

The absolute magnitudes of LHS~6328 in the near infra-red are consistently
$\sim$$0.8$\,mag fainter than the expected average value from the $8$\,pc sample
from Table~3 of \citet{2014A&A...569A.120L}. When compared with the average
photometry from \citet{2013ApJS..208....9P}\footnote{\url{http://www.pas.rochester.edu/~emamajek/EEM_dwarf_UBVIJHK_colors_Teff.txt}}, this M dwarf is also
consistently fainter in the optical by the same amount. However, it is worth
noting that $\sim$$0.9$\,mag is within the $2$ standard deviation of the
sample used for averaging in \citet{2014A&A...569A.120L}. This brightness
also coincidentally puts it very close to the MS gap in the optical,
discovered from the $100$\,pc sample from {\it Gaia} DR2~\citep{2018ApJ...861L..11J}.
See more discussion in Section~\ref{sec:outlier}.

\subsection{PSO~J1801+625}
\label{sec:ucwd}
The final calibrated spectra of the UCWD are shown in
Fig.~\ref{fig:UCWD_spectrum}. The usable spectral ranges are $3750-11000,
11500-13500, 14500-18000$ and $19500-24000$\,\AA. A black-body spectrum at
$3808$\,K is plotted as a visual reference. The combined spectrum is smooth
and does not show any absorption features, which could be the case for an
extreme subdwarf. The only absorption features present were identified as
coming from the residual sky lines, consistent with the spectrum expected
from a WD with a cool atmosphere. Catalogued photometry from Pan-STARRS~1
g$_{\mathrm{P1}}$, r$_{\mathrm{P1}}$, i$_{\mathrm{P1}}$ and z$_{\mathrm{P1}}$,
GDR2 G, G$_{\mathrm{BP}}$ and G$_{\mathrm{RP}}$ and ensemble
photometry with the field sources from the Liverpool Telescope $g$', $r$', $i$',
$z$' and H$_{\mathrm{LT}}$, and from the GTC J$_{\mathrm{EMIR}}$,
H$_{\mathrm{EMIR}}$ \& K$_{\mathrm{EMIR}}$ photometry that were converted
to flux values using the respective pivot wavelengths and overplotted onto
the spectra. The photometric SED shows a remarkable resemblance to that of
LSR~J0745+2627 if placed at the same distance~\citep{2012A&A...546L...3C}.

\subsection*{Model atmosphere analysis}
We performed a fit to the available spectra of the UCWD to extract its
atmospheric parameters. To do so we used the atmosphere models described
in \cite{blouin2018a,blouin2018b}. Different atmospheric compositions
were tested (pure helium, pure hydrogen and mixed compositions with H/He
abundance ratios ranging from $10^{-5}$ to $1$) and we found that the best
solution is by far a pure hydrogen model. Only a pure hydrogen model can
successfully reproduce both the IR flux depletion due to H$_2-$H$_2$
collision-induced absorption~\citep[CIA,][]{borysow2001} and the flux deficit
at small wavelengths due to the far red wing of the Lyman $\alpha$ line
broadened by collisions with H and H$_2$~\citep{kowalski2006}. If helium is
added to the model, the photospheric density becomes much higher, where
H$_2-$He CIA dominates~\citep{blouin2017} and the model underestimates the IR
flux.

The effective temperature and the solid angle were obtained by adjusting the
model fluxes to the observed spectra using a Levenberg--Marquardt algorithm.
As the distance $D$, $72.962$\,pc, is known from the {\it Gaia} parallax, the
radius $R$ can be computed from the solid angle $\pi R^2 / D^2$. Then, thanks
to evolutionary models, the radius and the effective temperature can be used
to obtain the mass (and $\log g$) and the cooling time of the white dwarf.
As the surface gravity resulting from this calculation was different from the
one initially assumed to perform the fit, we repeated the procedure described
above until the surface gravity converged.

\begin{figure}
\includegraphics[width=\columnwidth]{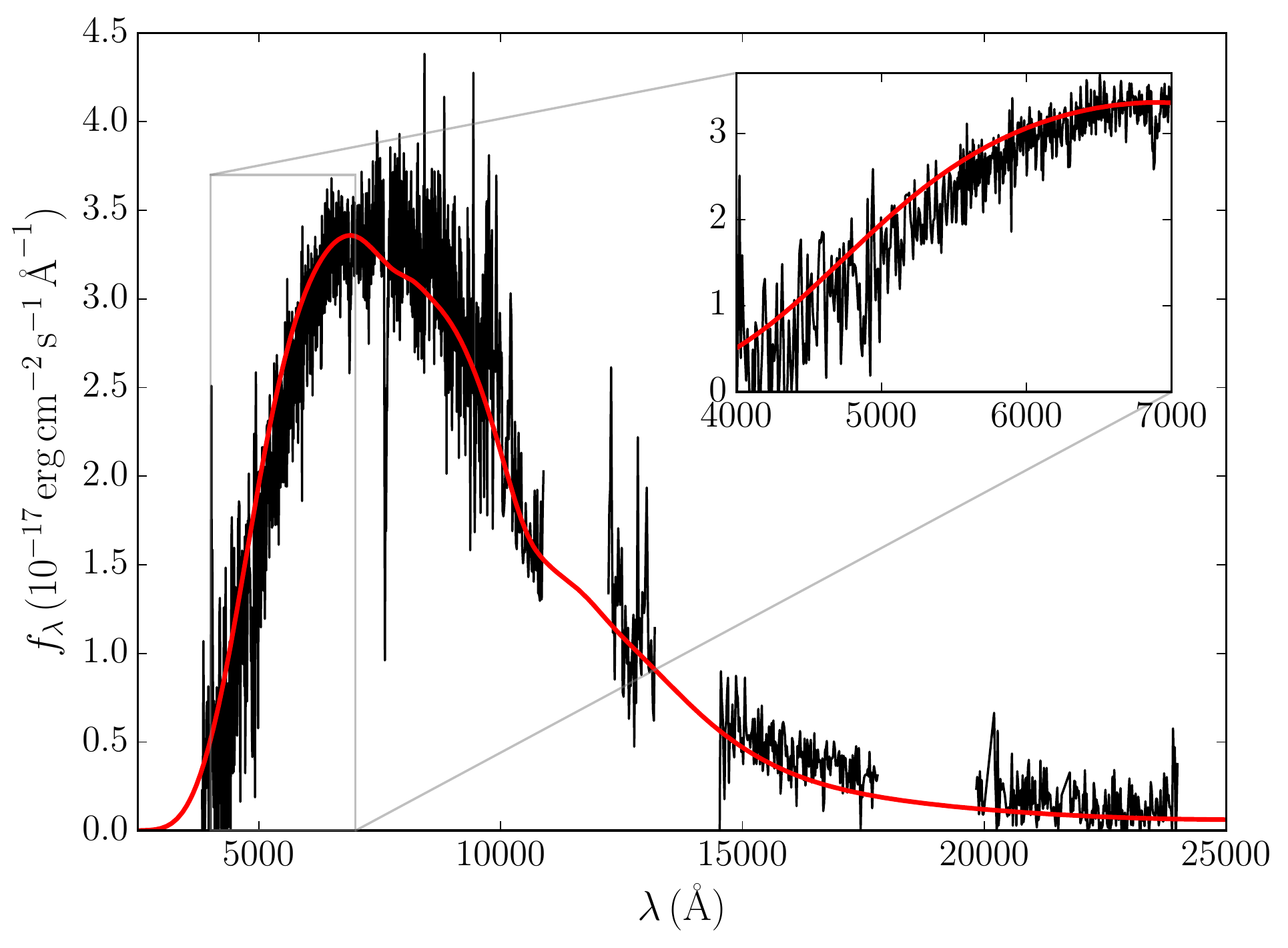}
\caption{The best fit model of PSO~J1801+625 indicates an effective temperature of $3550$\,K with low surface gravity. The best fit parameters are given in Table~\ref{tab:WDparameters}. The inset shows the small offset described in text.}
\label{fig:WDfit}
\end{figure}

Our best model achieves a good match to the whole spectral energy
distribution~(SED) of PSO~J1801+625 as shown in Fig.~\ref{fig:WDfit}. There
seems, however, to be an offset between the model and the observations
between $4000$ and $6000$\,{\AA} that appears to be due to an underestimation
of the strength of the Lyman $\alpha$ broadening~(see inset of Fig.~\ref{fig:WDfit}). The atmospheric parameters
of our best fit model are given in Table~\ref{tab:WDparameters}. Note that
the effective temperature is similar to that of LSR~J0745+2627~\citep[$T_{\rm eff}=3880 \pm 90\,{\rm K}$,][]{catalan2012}, unsurprising given the 
similarity between the SEDs of both objects (see Fig.~\ref{fig:UCWD_spectrum}). In
Table~\ref{tab:WDparameters}, we give the mass and cooling
time corresponding to the case where the white dwarf has a CO core and to
the case where it has a He core (which is more appropriate given the low mass
found).  For the CO core case we rely on the models of
\cite{fontaine2001} with $q({\rm He}) \equiv M_{\rm He} / M_{\star} = 10^{-2}$
and a thick hydrogen envelope of $M_{\rm H} / M_{\star} = 10^{-4}$.
As for the He core case, we use the evolutionary tracks of 
\cite{althaus2001}\footnote{\url{http://fcaglp.fcaglp.unlp.edu.ar/~althaus/}},
assuming once again $M_{\rm H} / M_{\star} = 10^{-4}$.

\begin{table}
    \centering
    \begin{tabular}{lcc}
                                & CO core & He core \\\hline\hline
        $T_{\rm eff}$ (K)       & \multicolumn{2}{c}{$3550 \pm 150$} \\
        He/H                    & \multicolumn{2}{c}{$0$} \\
        $R$ ($R_{\sun}$)        & \multicolumn{2}{c}{$0.0171 \pm 0.0012$} \\
        $M$ ($M_{\sun}$)        & $0.30 \pm 0.05$ & $0.33 \pm 0.06$ \\
        $\log g$                & $7.45 \pm 0.13$ & $7.49 \pm 0.13$ \\
        $\tau_{\rm cool}$ (Gyr) & $5.1 \pm 0.6$   & $10.5 \pm 1.5$ \\\hline
    \end{tabular}
    \caption{Atmospheric parameters of PSO~J1801+625.}
    \label{tab:WDparameters}
\end{table}

\section{Discussion}
To date, there are, including this work, $19$ field UCWDs with their
spectra confirmed. Their rarity and the lack of line features have
made their investigation difficult at best. The lack of parallax
measurements prior to {\it Gaia} DR2 and the inhomogeneity in the
sample characteristics makes it difficult to do rigorous statistical
analysis. Seven of the identified UCWDs were assumed to have a typical
white dwarf surface gravity $\log{g}=8.0$~\citep{2010ApJS..190...77K,
2016MNRAS.462.2295H}; while the remaining $11$ had the gravity
fitted as a free parameter: SSSJ1556-0806~\citep{2008MNRAS.385L..23R},
J0146+1404, J1001+3903, J1238+3502, J1251+4403, J2239+0018A, 
J2242+0048,~\citep{2015MNRAS.449.3966G},
LSH~3250~\citep{2015ApJS..219...19L},
WD~0346+246~\citep{2012MNRAS.423L.132K},
LHS~342, WD~0205-053~\citep{2019ApJ...878...63B}
\footnote{All the references here only refer to the most recently
analysis that uses the most updated atmosphere models, most of them are
not the original discovery articles.}.
What is most surprising is how $9$
of these $11$ UCWDs have $\log{g}<8.0$, the average mass of these being
$0.36$\,M$_{\sun}$. The least massive one, J2239+0018A, has
$\log{g}=6.95$, corresponding to a mass of $0.2$\,M$_{\sun}$. 
For the two non-low mass UCWDs, WD\,0346+246 and SSSJ1556-0806,
both of them have their distances underestimated as compared to the
Gaia distance from \citet{2018AJ....156...58B}. Due to the
degenerate dependency on the radius and distance of the object, when
an object is fitted with a distance that is too small, the radius
will also be too small.
Hence, these two UCWDs, having $\log{g}=8.3$ and $8.0$,
respectively, are almost certainly going to be low-mass. Though, a
thorough re-analysis is necessary to perform rigorous statistical
analysis over the whole sample of UCWDs before drawing any 
conclusions.

We note that an over-density of low-mass WDs is expected at low
effective temperatures as the result of the slowdown in their cooling
due to latent heat release from the core crystallization and the
extra thermal energy resulting from the first phase of convective
coupling~(see Fig.~15 of \citealp{2019ApJ...876...67B}). However,
stellar evolution models show that it takes longer than the age of
the Universe for an isolated star with M$<0.45$\,M$_{\sun}$ to turn
into a WD. In this sense, the $0.36$\,M$_{\sun}$ average mass for
UCWDs is puzzling.

\subsection{Low Mass Ultra-Cool White Dwarf}
We have arrived with a few possibilities that can be attributed to the bulk
of the low mass UCWDs in the known sample: (1)~ higher mass ones do not
exist; (2)~unaccounted input physics in the current UCWD model, supported by
the poor fits found in \citet{2015MNRAS.449.3966G}; (3)~they belong
to a population of WD that evolved through a specific pathway.

\subsection*{End State of Single Metal-rich Star}
There is evidence that an isolated metal-rich ($0.3<[$Fe$/$H$]< 0.5$;
\citealt{2006ApJ...646..499O, 2006ApJ...642..462G}) star can lose
sufficient mass to form a low mass WD~\citep{2005ApJ...635..522H,
2007ApJ...671..748K, 2007ApJ...671..761K}.  In order to become an UCWD,
they must have cooled for a few billion years regardless of the formation
scenario. It would seem rather unlikely that the majority of the $11$ low
mass UCWDs have high metallicity. Furthermore, with the current picture
of metal enrichment process of the Milky Way, it is unlikely to have this
many old stars with enhanced metallicity in the early time of the
Galaxy.

\subsection*{Common Envelope Evolution}
The formation of a low-mass WD requires significant
mass loss in the post-MS stage. The current understanding is that
through common envelope~(CE) evolution, the binary will produce a low
mass WD; while in a stable Roche lobe overflow system, an extremely low
mass WD would be produced instead~\citep{2019ApJ...871..148L}. Post-CE
systems are tight binaries~(hard), dynamically hard binaries become
harder~(separation decreases) when interacting with a third
body~(known as the Heggie's Law, see \citealt{1975MNRAS.173..729H}). If
they were formed in binary systems, where is the missing companion star?
Given the faintness of the UCWD, typically with M$_{\mathrm{bol}}>15$,
any companions earlier than mid-T dwarf would have shown up in the
optical and near-infrared images and/or spectra~(see also the next
scenario). It should be noted that given the low mass of a brown dwarf,
if it should exist in the progenitor system, it could not have played a
significant role in the late stage of the stellar evolution of the WD
progenitor.

\subsection*{Unseen Companion(s)}
\label{sec:unseen_companion}
Using the IRSA\footnote{\url{https://irsa.ipac.caltech.edu/}} catalogue and
time series tool, ZTF DR1~\citep{2019PASP..131a8002B, 2019PASP..131f8003B,
2019PASP..131g8001G} with over $50$ epochs in g and r bands, no variability
is observed by either the UCWD or the M dwarf. Cross-correlations of the
$3$ up-sampled medium resolution spectra show shifts of $0.022$ pixels
between the first and second epoch and $-0.004$ pixels between the second
and third epoch. This corresponds to $0.6$ and $-0.1$\,km\,s$^{-1}$
respectively --- with signal-to-noise ratios under $0.1$, such that we
disregard the values. Hence, it does not support a close, unresolved
companion of any significant mass. The $Ks$ band photometry reached
$\sim$$21$\,mag. At $\sim$$70$\,pc a mid-T dwarf should be
detectable~\citep{2012ApJ...752...56F, 2018MNRAS.481.5216C}, which can
be in the form of IR excess if they are unresolved. Similarly, if they
are resolved, they would be bright enough to show up in the W2 image at
$70$\,pc~\citep{2014ApJ...796...39T}.

Could they be the end products of low-mass X-ray binaries with a hidden
neutron star in this system? \citet{2001ASSL..264..355P} suggests that
the end products are predominantly He and low-mass hybrid
HeCO WDs. When the pulsar beam is not pointing towards us, there would
not be any direct detection of the neutron star. Without absorption lines
with which to measure radial velocity it is not possible to infer the
presence of such a compact heavy object. Transits are not present in most
binary systems, due to orbital inclination. In the stable
Roche-lobe overflow or double helium core WD merger formation
scenario of hot subdwarf B stars, the low-mass population cannot burn
helium and they eventually evolve to become helium core
WDs~\citep{2002MNRAS.336..449H,2003MNRAS.341..669H}.
This tends to produce low mass WDs. In the case of both He and CO
core WDs, this can be possible for symbiotic stellar evolution that happened
at the early times of the Milky Way. HD~188112 is the prototype that
bridges the late hot subdwarf~B to a He core WD~\citep{2016A&A...585A.115L}. 

\subsection{Typical Mass Ultra-Cool White Dwarfs}
All empirical or theoretical initial-final mass 
relations~\citep{2008A&A...477..213C, 2012ApJ...746..144Z,
2015ApJ...815...63A, 2018ApJ...860L..17E} tell the same story:
\textit{more massive stars leave behind more massive remnants}.
UCWDs are old objects, they have been cooling for a few billion years.
From the PARSEC stellar evolution model~\citep{2012MNRAS.427..127B},
$1.5$ and $2$M$_{\sun}$ solar-metallicity stars take $\sim$$3$ and
$\sim$$1.5$\,Gyr to become WDs, respectively. Their corresponding WD
masses are $\sim$$0.6$M$_{\sun}$, which would take $\sim$$10$\,Gyr for
them to cool to $3500$\,K, assuming a pure hydrogen atmosphere. The
combined age of  $11-13$\,Gyr would be roughly that of the thick disc
and the halo. For a $\sim$$0.9$M$_{\sun}$ WD, which would have been a
$\sim$$5$M$_{\sun}$ MS star. Due to a smaller radius and low
luminosity, it would take $\sim$$11$\,Gyr to reach $3500$\,K~(see
Fig.~12 of \citealp{2019ApJ...876...67B}, and note that
$\sim$$0.8$M$_{\sun}$ WD has the lowest cooling rate). More massive
stars evolve quicker, so the total time required would be roughly the
same as that required by WDs with typical masses.

A metal poor star evolves faster, so it is possible for a MS star
slightly under $1.5$M$_{\sun}$ to become WDs, however, they
also have lower total mass loss resulting in slightly more massive
WDs that cool slower. Therefore, the total MS life time and WD cooling
time for the metal poor systems is similar to those with solar
metallicity.

The seven remaining spectroscopically confirmed UCWDs in the
Montreal White Dwarf
Database\footnote{\url{http://www.montrealwhitedwarfdatabase.org}}
do not have the surface gravity fitted as a free parameter, so
it is unclear if any of those would fit into this category of
UCWDs that we can treat them as singly evolved stars.

In order for PSO~J1801+625 to be a typical mass WD, it has to be
$\sim$$45\%$ closer. This requires the parallax to be $10\sigma$
off for the UCWD and a few hundreds for the equidistant M dwarf.
On the other hand, if the ``low mass UCWD'' is a pair of unresolved
binary UCWDs, because of the degenerate nature of WD, the more
massive it is, the smaller it becomes. At a given distance and
temperature, their radii is a factor of $\sqrt{2}$ smaller in 
order for them to have the same luminosity as a single low mass
WD. This decrease would bring the mass of the WDs up to that of a
typical WD. The total life time of the ``unresolved binary
CO-WDs'' would be too long compared that expected for a CO-core
WD. In any case, it  does not seem probable to have all or most
of the UCWDs mentioned in the previous section to be unresolved
binaries with almost identical mass, radius, envelope composition
and temperature.

\subsection{What if all our measurements are outlying data points?}
\label{sec:outlier}
For a solar metallicity single M1.75 dwarf to have the apparent
magnitude as we observed, it should be at $\sim$$110$\,pc. The
discrepancy in the absolute magnitude of LHS~6328 and that from the
averaged values of the solar neighbourhood sample is roughly
$0.8$\,mag in all optical and near-infrared filters.
NEOWISE~\citep{2010AJ....140.1868W, 2014AAS...22321708M} has over
$1000$ epochs in W1 \& W2 bands, where they do not show sign of
variability using the periodogram tool available on their webpage.
Data from the Transiting Exoplanet Survey
Satellite~\citep[TESS,][]{2015JATIS...1a4003R, 2018SPIE10704E..15S}
were period folded with \texttt{Period04}~\citep{2005CoAst.146...53L}
between $0$ and $360$ cycles/day in step of 0.000625 with no
$5\sigma$ peaks detected above 
$\sim$$0.4$\,mmag\footnote{TESS unique object ID: TIC 233068267}.
Hence, neither starspots nor occultation can be the explanation;
any object that is big enough to remove $0.8$\,mag of light would have
shown up in the spectra or photometry. If we consider any single
observable to see how much it takes to compensate for the
``lost flux'', it is not possible to explain the discrepancy
without leading to contradicting spectral, photometric and
astrometric information. In the following, we list the possibilities
to explain, if our data gathering and template fittings have been
unlucky throughout, and LHS~6328 is a mild outlier.
\begin{enumerate}
    \item In the {\it Gaia} DR2, LHS~6328 has an
        \textsc{astrometric\_excess\_noise\_sig} of
        $6.85$, meaning its astrometric solution is unreliable. For
        the WD, it is close to the detection limit, so the parallax
        is also unreliable. If LHS~6328 is at the upper distance
        limit, $73.1$\,pc, it woud lead to a dimming of $0.01$\,mag
        in the absolute magnitude.
    \item The uncertainty in our M dwarf template fitting is $0.25$
        spectral type, while the absolute error can be as large as
        $0.5$ spectral type, this can contribute a $0.1$\,mag of
        difference.
    \item The metallicity indicator $\zeta_{\mathrm{TiO/CaH}}$
        indicates a slightly sub-solar metallicity, which puts
        LHS~6328 close to the border of dwarf and subdwarf.
        A typical subdwarf has a metallicity of $[$Fe$/$H$]=-0.5$;
        if LHS~6328 has $[$Fe$/$H$]=-0.25$, and combined with a
        shift in $+0.25$ spectral type~(i.e.\ lower temperature),
        due to the reduced radiative pressure in the atmosphere.
        The radius becomes smaller by about
        $20\%$~\citep{1997A&A...327.1054B, 2019AJ....157...63K}.
        This would correspond to a change of $0.2$\,mag.
    \item In the case of SDSS filters, the dispersion in the
        magnitudes for a given spectral type is about $0.25$\,mag~\citep{2005PASP..117..706W,2010AJ....139.1808S}.
\end{enumerate}
If LHS~6328 is $1.5\sigma$ from the average in all quantities,
the total magnitude difference could reach $0.8$\,mag. This 
can account for the slightly spurious luminosity of a system
containing two single objects at $\sim$$70$\,pc, as supported by
the Renormalised Unit Weight Error (RUWE, see
Table~\ref{tab:observed_properties}) which are close to $1$.

\section{Conclusion}
We have collected the spectra of an UCWD in common proper
motion with a slightly metal poor M1.75 dwarf star, LHS\,6328. The
UCWD has an effective temperature of $3550$\,K, given its low mass
at $0.33$\,$M_{\sun}$, it is likely to have a He core, with a total
cooling age of $10.5\pm1.5$\,Gyr. This is most
likely the remnant from a previously interacting binary that the
companion is no where to be seen. This is the first system of a
field UCWD with a pure hydrogen atmosphere that we can obtain the
progenitor metallicity, $-0.25<[$Fe$/$H$]<0$, from its coeval main
sequence companion. From the radial velocity measurement of the
LHS~6328, we also have the first full 6D kinematics found for a UCWD,
with $\left(U, V, W\right) = \left(-114.26\pm0.24, 222.94\pm0.60, 10.25\pm0.34\right)$\,km\,s$^{-1}$, suggesting a thick disc-like
velocity. Despite these measurement, the unclear origin of the low
mass UCWD does not allow us to confidently draw any concrete
conclusions. We propose a number of explanations but there is no
evidence for any of the suggested scenarios. We believe it is
necessary to run simulations on a wide range of configurations in
order to understand what kind of evolution scenarios could have led
to a system like LHS~6328 and PSO~J1801+625.

\section*{Acknowledgements}
ML acknowledges financial support from the OPTICON. ML thanks P. R. McWhirter
for the frequency analysis on the M-dwarf variability.

NL was financially supported by the Spanish Ministry of Economy and
Competitiveness (MINECO) under the grant AYA2015-69350-C3-2-P (2016-2019). NL thanks Lee Patrick, Alina Streblyanska, and the pyemir developers for their help during the reduction of the EMIR data.

SB acknowledges support from the Laboratory Directed Research and Development program of Los Alamos National Laboratory under project number 20190624PRD2.

The Liverpool Telescope is operated on the island of La Palma by
Liverpool John Moores University in the Spanish Observatorio del
Roque de los Muchachos of the Instituto de Astrof\'isica de Canarias
with financial support from the UK Science and Technology Facilities
Council.

The William Herschel Telescope and its service programme are operated
on the island of La Palma by the Isaac Newton Group of Telescopes in
the Spanish Observatorio del Roque de los Muchachos of the Instituto
de Astrof\'isica de Canarias.

Based on observations made with the Gran Telescopio Canarias~(GTC), installed 
in the Spanish Observatorio del Roque de los Muchachos of the Instituto de 
Astrof\'isica de Canarias, in the island of La Palma.

This work is partly based on data obtained with the instrument EMIR, built by a Consortium led by the Instituto de Astrof\'isica de Canarias. EMIR was funded by GRANTECAN and the National Plan of Astronomy and Astrophysics of the Spanish Government.

This publication makes use of data products from the Two Micron All Sky Survey, which is a joint project of the University of Massachusetts and the Infrared Processing and Analysis Center/California Institute of Technology, funded by the National Aeronautics and Space Administration and the National Science Foundation.

This publication makes use of data products from the Wide-field Infrared Survey Explorer, which is a joint project of the University of California, Los Angeles, and the Jet Propulsion Laboratory/California Institute of Technology, funded by the National Aeronautics and Space Administration.

This paper includes data collected by the TESS mission. Funding for the TESS mission is provided by the NASA Explorer Program.

The Pan-STARRS1 Surveys (PS1) and the PS1 public science archive have been made possible through contributions by the Institute for Astronomy, the University of Hawaii, the Pan-STARRS Project Office, the Max-Planck Society and its participating institutes, the Max Planck Institute for Astronomy, Heidelberg and the Max Planck Institute for Extraterrestrial Physics, Garching, The Johns Hopkins University, Durham University, the University of Edinburgh, the Queen's University Belfast, the Harvard-Smithsonian Center for Astrophysics, the Las Cumbres Observatory Global Telescope Network Incorporated, the National Central University of Taiwan, the Space Telescope Science Institute, the National Aeronautics and Space Administration under Grant No. NNX08AR22G issued through the Planetary Science Division of the NASA Science Mission Directorate, the National Science Foundation Grant No. AST-1238877, the University of Maryland, Eotvos Lorand University (ELTE), the Los Alamos National Laboratory, and the Gordon and Betty Moore Foundation.

This work has made use of data from the European Space Agency (ESA) mission
{\it Gaia} (\url{https://www.cosmos.esa.int/gaia}), processed by the {\it Gaia}
Data Processing and Analysis Consortium (DPAC,
\url{https://www.cosmos.esa.int/web/gaia/dpac/consortium}). Funding for the DPAC
has been provided by national institutions, in particular the institutions
participating in the {\it Gaia} Multilateral Agreement.




%%%%%%%%%%%%%%%%%%%%%%%%%%%%%%%%%%%%%%%%%%%%%%%%%%



\bibliographystyle{mnras}
\bibliography{2019UCWD} % if your bibtex file is called example.bib


%%%%%%%%%%%%%%%%%%%%%%%%%%%%%%%%%%%%%%%%%%%%%%%%%%

%%%%%%%%%%%%%%%%% APPENDICES %%%%%%%%%%%%%%%%%%%%%

\appendix

\section{Appendix}
The 8 lines used for computing the radial velocity of the M dwarf are listed in Table~\ref{tab:rv_lines}. The spectrum used for the measurement is the median of the three 20-minutes FRODOSpec spectra.
\begin{table}
    \centering
    \begin{tabular}{c|c|c}
        BT-SETTL / {\AA} & Measured / {\AA} & Radial Velocity / km\,s$^{-1}$\\\hline\hline
        6122.68 & 6122.5 & -0.88\\
        6162.63 & 6162.5 & -0.63\\
        6450.30 & 6449.9 & -1.86\\
        6463.06 & 6463.0 & -0.28\\
        7326.70 & 7326.4 & -1.23\\
        7511.59 & 7511.3 & -1.16\\
        7665.50 & 7665.1 & -1.57\\
        7699.56 & 7699.0 & -2.18\\\hline\hline
    \end{tabular}
    \caption{The wavelengths of the model absorption lines from BT-SETTL are corrected with an air density of $0.733$\,amg based on the observing conditions. The $8$ lines that can be identified by eyes are used to compute the radial velocities.}
    \label{tab:rv_lines}
\end{table}


%%%%%%%%%%%%%%%%%%%%%%%%%%%%%%%%%%%%%%%%%%%%%%%%%%

% Don't change these lines
%\bsp	% typesetting comment
\label{lastpage}
\end{document}